\documentclass{PoS}

\usepackage{subfigure,amsmath}

\newcommand{\comments}[1]{}

\title{Holographic study of rho meson mass in an external magnetic field\\
{\normalsize Paving the road towards a magnetically induced superconducting
QCD vacuum?}}

\ShortTitle{Holographic study of rho meson mass in an external magnetic
field}

\comments{
\author{\speaker{Nele Callebaut}

         \thanks{A footnote may follow.}\\

        UGent, Belgium\\

        E-mail: \email{ncalleba.Callebaut@UGent.be}}
}

\author{\speaker{Nele Callebaut}, David Dudal, Henri Verschelde \\
        Ghent University, Department of Physics and Astronomy \\
        Krijgslaan 281-S9, 9000 Gent, Belgium\\
        E-mail: \email{ncalleba.callebaut@UGent.be} ,\email{
david.dudal@ugent.be},\email{ henri.verschelde@ugent.be}}

\abstract{We study the rho meson mass in a uniform
background magnetic field $eB$ at zero temperature, in search of indications
for the magnetically induced rho meson condensation, as predicted recently
by Chernodub.
The holographic model used is the Sakai-Sugimoto model with two flavours and
a non-zero constituent quark mass. We fix the free holographic parameters by
matching them to the phenomenological value for the constituent quark mass
and the experimental values for
the pion decay constant and the rho meson mass, this in absence of a
magnetic field. In a first approximation, the Landau levels are recovered,
indeed indicating an instability of the QCD vacuum at a critical magnetic
field, $eB_{c} \sim m_\rho^2$, to a phase where rho mesons are condensed. We
improve on this result by also taking into account the holographic analogue of chiral magnetic catalysis, numerically solving the mass eigenvalue
equation for the rho meson, which depends on $eB$ both explicitly and implicitly through the
changed embedding of the flavour probe branes. This turns out to raise
$eB_{c}$ with a few percents. As a byproduct of our analysis we find that
the separation between the chiral symmetry restoration temperature
$T_\chi(eB)$ and the deconfinement temperature $T_{c}$ is 3.2 percent at $eB
= 30 m_\pi^2 \approx 0.57$ GeV$^2$.    }

\FullConference{The many faces of QCD\\

		 November 1-5, 2010\\

		 Gent, Belgium}

\begin{document}

\section{Introduction}

The interest in QCD in the presence of a
strong magnetic background has arisen recently since strong magnetic fields are expected to appear in heavy ion collisions \cite{Skokov:2009qp}.
In \cite{Chernodub:2010qx,Chernodub:2011mc} it was suggested that the cold
$(T=0)$ QCD vacuum is unstable towards formation of a condensate of charged
rho mesons when a sufficiently strong external magnetic field\footnote{For
notational convenience, we shall usually write just $B$, rather than $eB$.}
($B \sim m_\rho^2$) is applied.
The argumentation in \cite{Chernodub:2010qx} is based
on the DSGS-Lagrangian \cite{Djukanovic:2005ag} describing self-consistent
quantum electrodynamics for the rho mesons, whereas in
\cite{Chernodub:2011mc} a similar reasoning was built up in a NJL-model.

Inspired by these papers and the already known existence of rho meson
condensation at finite isospin chemical potential in holographic QCD
\cite{Aharony:2007uu}, we have studied the dependence of the rho meson mass
on a background magnetic field in a holographic approach. Holographic
QCD-models give a description of hadronic physics through a dual
supergravity theory in a higher-dimensional world. The duality is valid at
large 't Hooft coupling where QCD itself is unmanageable, thus providing a
setting for studying non-perturbative QCD effects. The holographic model
which we have used, the Sakai-Sugimoto model
\cite{Sakai:2004cn,Sakai:2005yt}, manages to reproduce much of the
low-energy physics of QCD, in particular confinement and dynamical chiral symmetry
breaking at low temperature.

In section \ref{Holographic} we give a short review of this model, including
our numerical values for the free parameters obtained by matching to
phenomenological and experimental results, and we describe the method to
introduce the magnetic field. The effect of the magnetic field on the
geometry of the model in the current approximation is summarized in section
\ref{Influence}, and then used in section  \ref{EOM} to determine the
$B$-dependence of the rho meson mass and the critical value of the magnetic
field at which the charged rho mesons condense. In the last section we
calculate the magnetically induced split between deconfinement and chiral
symmetry restoration temperature, finding a reasonable agreement with
phenomenological and lattice output.
In this proceeding, we only summarize our
findings, details will appear shortly in \cite{nele}.

\section{Holographic setup} \label{Holographic}

The Sakai-Sugimoto model \cite{Sakai:2004cn,Sakai:2005yt} involves a system
of $N_f$ pairs of D8-$\overline{\mbox{D8}}$ flavour probe branes in the
D4-brane background
\[
ds^2 = \left(\frac{u}{R}\right)^{3/2} (\eta_{\mu\nu}dx^\mu dx^\nu + f(u)d\tau^2) + \left(\frac{R}{u}\right)^{3/2}
\left( \frac{du^2}{f(u)} + u^2 d\Omega_4^2 \right), \nonumber
\]
\[
e^\phi = g_s \left(\frac{u}{R}\right)^{3/4} \hspace{2mm}, \quad
F_4 = \frac{N_c}{V_4}\epsilon_4 \hspace{2mm}, \quad f(u) = 1-\frac{u_K^3}{u^3},
\]
\noindent where $d\Omega_4^2$, $\epsilon_4$ and $V_4=8\pi^2/3$ are,
respectively, the line element, the volume form and the volume of a unit
four-sphere,  while $R$ is a constant parameter related to the string
coupling constant $g_s$, the number of colours $N_c$ and the string
length $\ell_s$ through $R^3 = \pi g_s N_c \ell_s^3$.
This background has a
natural cut-off at $u = u_K$ and the QCD-like theory is said to ``live'' at
$u \rightarrow \infty$. We set $N_c = 3$, and the number of flavours $N_f=2$.
We stress here that we ignore the back reaction of
the flavour branes on the background, meaning we are working in a
holographic analogue of the quenched approximation.

The Sakai-Sugimoto model incorporates dynamical chiral symmetry breaking
as a consequence of the $U$-shaped embedding of the flavour branes: the
D8-branes and $\overline{\mbox{D8}}$-branes merge at a certain value of the
extra holographic dimension $u = u_0$. In contradistinction to the
original setup of \cite{Sakai:2004cn,Sakai:2005yt}, we consider the
more general setting with $u_0 > u_K$, this in order to have non-zero
constituent quark mass \cite{Aharony:2006da}
\begin{equation}\label{constmass}
    m_q=\frac{1}{2\pi \alpha'}\int_{u_K}^{u_0}\frac{du}{\sqrt{f(u)}}\,,
\end{equation}
with $\alpha'= \ell_s^2$ the string tension. The bare quark masses are
always zero, so we are working in the chiral limit.

The gauge field $A_m(x^\mu,u)$ ($m=0,1,2,3,u)$ living on the D8-branes
describes  mesons in the boundary field theory. The action for this gauge
field is given by the non-Abelian DBI-action\footnote{Because the 't Hooft
coupling $\lambda$ is assumedly large, we ignore in a first approximation
the Chern-Simons part of the action in the analysis, being a factor
$1/\lambda$ smaller than the DBI-action.}
\begin{equation} \label{S}
S_{DBI} = -T_8 \int d^4x d\tau \hspace{1mm}\epsilon_4 e^{-\phi} \textrm{STr}
\sqrt{-\det \left[g_{mn}^{D8} + (2\pi\alpha') F_{mn} \right]}
\end{equation}
with $T_8 = 1/((2\pi)^8 \ell_s^9)$ the D8-brane tension, $\textrm{STr}$ the
symmetrized trace,  $g_{mn}^{D8}$ the induced metric on the D8-branes, and
$F_{mn} = \partial_m A_n - \partial_n A_m + [A_m, A_n] = F_{mn}^a t^a$ the
field strength with anti-hermitian generators
\[ t^a  =  \frac{i}{2} (\textbf{1}, \sigma_1, \sigma_2, \sigma_3), \qquad
\textrm{Tr}(t^a t^b) = - \frac{\delta_{ab}}{2}. \]

\begin{figure}[t]
  \centering
  \scalebox{0.35}{
  \includegraphics{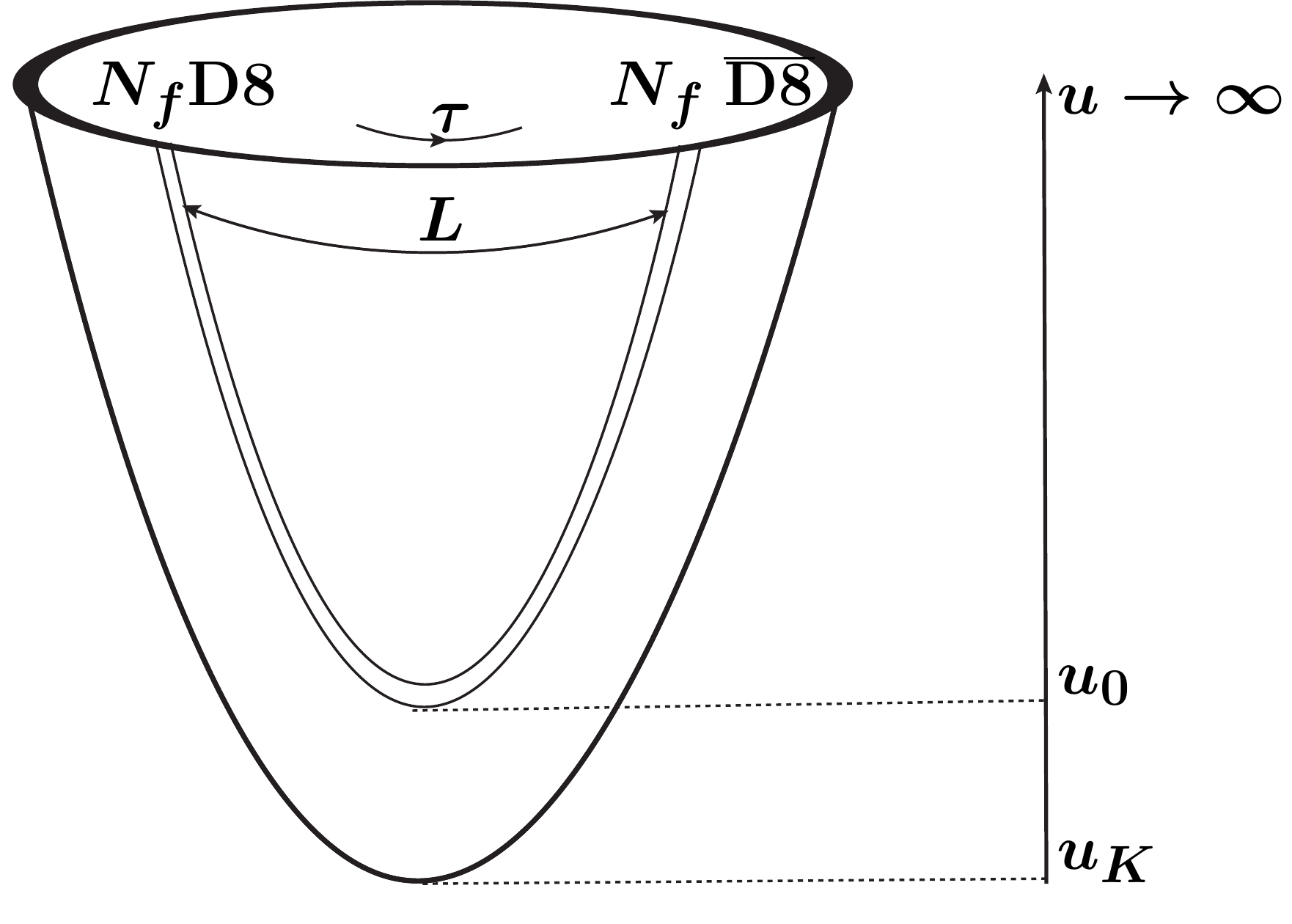}}
  \caption{The Sakai-Sugimoto model.}\label{finiteTfig1}
\end{figure}
The parameters $R, g_s , \ell_s, M_K$, $u_K$ and $\lambda = g^2_{YM} N_c$ are
related through the following equations:
\begin{equation} \label{relatie1}
R^3 = \frac{1}{2} \frac{\lambda \ell_s^2}{M_K},  \quad g_s =
\frac{1}{2\pi}\frac{g^2_{YM}}{M_K \ell_s},  \quad u_K = \frac{2}{9} \lambda
M_K \ell_s^2.
\end{equation}
Without loss of generality one can put $u_K = \frac{1}{M_K}$
\cite{Sakai:2005yt}. To fix the seven unknown parameters ($R$,
$\lambda$, $\ell_s$, $M_K$, $u_K$, $g_s$ and $u_0$) we thus need three additional
conditions. We intend to choose these in such a way that the major
features of $N_f=2$ QCD are reproduced as well as possible. In particular,
we use the phenomenological value for the constituent quark mass ($m_q
= 0.310$ GeV, as in \cite{Mizher:2010zb}), and the experimental values for
the pion decay constant ($f_\pi = 0.093$ GeV) and for the rho meson mass in
absence of magnetic field ($m_\rho = 0.776$ GeV). The results of our
numerical analysis are \cite{nele}
\begin{equation}
M_K \approx 0.721 \mbox{  GeV}, \quad  u_0 \approx 1.92 \mbox{  GeV}^{-1}
\mbox{  and   } \kappa  = \frac{\lambda N_c}{216 \pi^3} \approx 0.006778.
\end{equation}
From this, we do extract a relatively large 't Hooft coupling,
$\lambda\approx 15$. Similar values can be found in \cite{Sakai:2005yt},
however these were obtained in the limiting case $u_0=u_K$, corresponding to
zero constituent quark mass. In this setting it would be, as we shall see
shortly, impossible to model a magnetic catalysis effect, which is by now
generally accepted to occur in QCD.

To turn on an electromagnetic background field $A_\mu^{em}$ in the boundary
field theory we use the standard technique \cite{Sakai:2005yt} of putting
\begin{equation}
A_\mu(u \rightarrow \infty) = e Q_{em} A_\mu^{em} = e  \left(
\begin{array}{cc} 2/3 & 0 \\ 0 & -1/3 \end{array} \right) A_\mu^{em} = e
\left( \frac{1}{6} \textbf{1}_2 + \frac{1}{2} \sigma_3 \right)  A_\mu^{em}
\end{equation}
which, for the case of an external magnetic field in the $x_3$-direction in
the boundary field theory ($F_{12}^{em} = \partial_1 A^{em}_2 = B$), amounts
to setting
\begin{equation}
A_\mu(u \rightarrow \infty) = e Q_{em} x_1 B \delta_{\mu2}= \frac{ x_1 e B
\delta_{\mu2}}{3} \left( \frac{i\textbf{1}_2}{2} \right)   +  x_1 e B
\delta_{\mu2} \left( \frac{i\sigma_3}{2}  \right),
\end{equation}
or, adapting the notation $\overline A_\mu = A_\mu(u \rightarrow \infty)$:
\begin{equation}
\overline A_2^3 = x_1 e B \quad \mbox{   and   } \quad \overline A_2^0 =
\overline A_2^3 / 3.
\end{equation}
Because we want to look at the effect of the electromagnetic field on the
rho mesons, which transform in the adjoint of $U(N_f)_L \times U(N_f)_R$
and hence only couple to the isospin component $\overline{A}_\mu^3$ of the
electromagnetic field, we may ignore $\overline{A}_\mu^0$.

We work in the $A_u = 0$ gauge and fix the residual gauge symmetry by
working in a particular gauge where the total gauge field is given by
\cite{Sakai:2005yt} 
\begin{equation} \label{A}
A_\mu(x^\mu,u) = \overline A_\mu  + \sum_{n\geq1} V_{\mu,n} (x^\mu)
\psi_n(u),
\end{equation}
with $V_{\mu,n}(x^\mu)$ a tower of vector mesons with masses $m_n$, and
$\{\psi_n(u)\}_{n\geq 1}$ a complete set of functions of $u$, satisfying the
eigenvalue equation
\begin{equation} \label{eigwvgl}
u^{1/2} \gamma_B^{-1/2}(u) \partial_u \left[ u^{5/2} \gamma_B^{-1/2}(u)
\partial_u \psi_n(u) \right] = -R^3 m_n^2 \psi_n(u),
\end{equation}
and the normalization condition
\begin{equation}
\frac{1}{2g_s} (2\pi \alpha')2 R^{3/2} T_8 V_4 \int_{u_0}^\infty du
\hspace{1mm} u^{-1/2} \gamma_B^{1/2} R^3 \psi_m(u) \psi_n(u) = \frac{1}{4}
\delta_{mn},
\end{equation}
with
\begin{equation}
\gamma_B(u) = \frac{u^8 A(u)}{u^8 A(u) f(u) - u_0^8 A_0 f_0 }\hspace{1mm},
\quad A(u) = 1+B^2 \left(\frac{R}{u}\right)^3, \quad A_0 = 1+B^2
\left(\frac{R}{u_0}\right)^3.
\end{equation}

\section{Influence of the magnetic field on the embedding of the flavour probe
branes: magnetic catalysis of chiral symmetry breaking} \label{Influence}

\begin{figure}[h]
  \hfill
  \begin{minipage}[t]{.45\textwidth}
    \begin{center}
      \scalebox{0.8}{
  \includegraphics{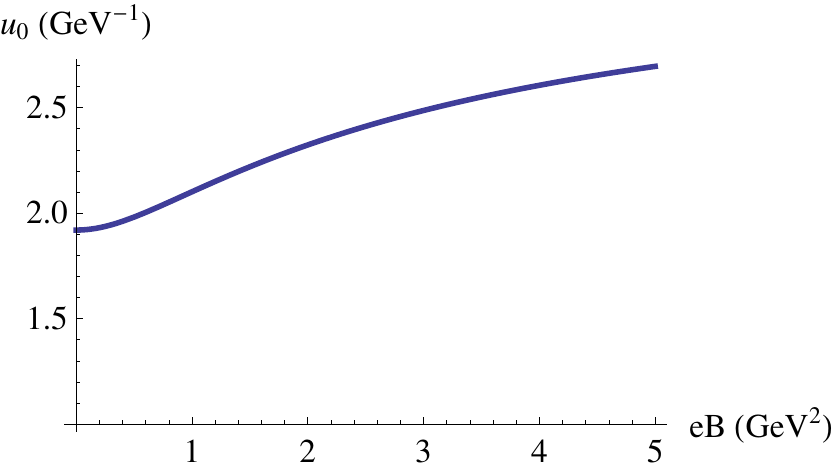}}
      \caption{$u_0$ as a function of the magnetic field.}
      \label{fig-tc}
    \end{center}
  \end{minipage}
  \hfill
  \begin{minipage}[t]{.45\textwidth}
    \begin{center}
      \scalebox{0.8}{
  \includegraphics{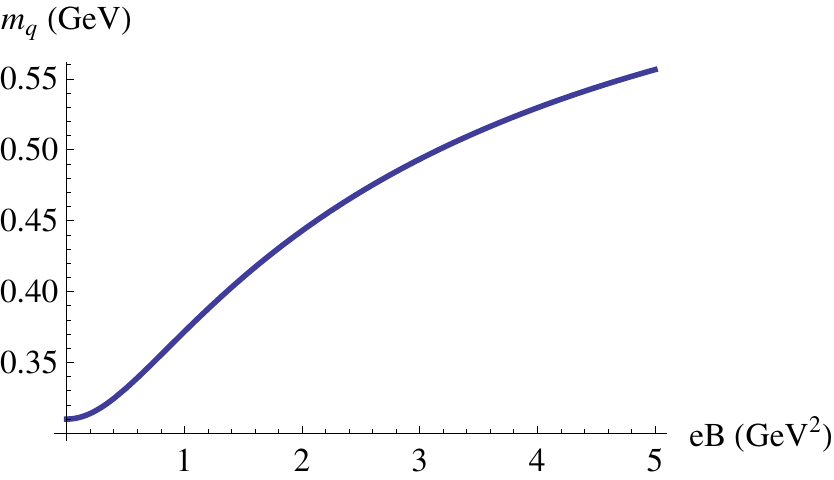}}
      \caption{The constituent quark mass as a function of the magnetic
field.}
      \label{mq}
    \end{center}
  \end{minipage}
  \hfill
\end{figure}

As already discussed in \cite{Johnson:2008vna} for the Sakai-Sugimoto model,
an external magnetic field promotes chiral symmetry breaking: when holding
the asymptotic separation $L$ between D8- and $\overline{\mbox{D8}}$-branes
fixed, $u_0$, which is in a one-to-one correspondence with $L$ \cite{Sakai:2004cn,Sakai:2005yt,Aharony:2006da},
grows with $B$.
This means that the probe branes get more and more bent
towards each other, driving them further and further away from the chirally
invariant situation (straight branes). This feature corresponds to a
holographic modeling of the magnetic catalysis of chiral symmetry breaking
\cite{Miransky:2002rp}. For our value of $L = 1.57$
GeV$^{-1}$ (corresponding to $u_0(B=0) = 1.92$ GeV$^{-1}$) kept fixed, the
numerically obtained $B$-dependence of $u_0$ and consequently of the
constituent quark mass $m_q$, via the relation (\ref{constmass}), is plotted
in Figure \ref{mq} \cite{nele}. Since $m_q$ rises with $B$, we expect that
taking this magnetic chiral catalysis into account will translate into
the rho meson mass $m_\rho$ also rising with $B$, at least when ignoring the lowest Landau level shift (see next section).

\section{Equations of motion for the mesons in the magnetic field}
\label{EOM}

Plugging the gauge field ansatz (\ref{A}) into the action (\ref{S}) and
expanding it to order $V_{\mu,n}^2$, we derive the equations of motion for
the vector mesons $V_{\mu,n}$, and thus in particular the rho mesons
$\rho_\mu$ (we concentrate on the lightest vector mesons, because they will
probably be the first to condense at strong $B$).
We find \cite{nele}
\begin{equation}
m_\rho^2 \rho_\mu^- - D_\nu^2 \rho_\mu^- + 2i G_{\mu\nu}\rho^{\nu -} = 0,
\quad \mbox{with  } D_\mu = \partial_\mu - ie A_\mu^{em},  \quad G_{\mu\nu}
= \partial_\mu A_\nu^{em} - \partial_\nu A_\mu^{em},
\end{equation}
for the negatively charged rho meson $\rho_\mu^- = \rho_\mu^1 -
i\rho_\mu^2$,
and the complex conjugate of this equation for the positively charged rho
meson $\rho_\mu^+ = \rho_\mu^1 + i\rho_\mu^2$.
Inserting the background gauge field ansatz $\overline{A}_2^3 = x_1 e B$
into the above equation of motion, and Fourier transforming $\rho_\mu^-$,
we obtain the Landau energy levels \cite{Nielsen:1978rm}
\begin{equation}  \label{landau}
E^2 (\rho_1^- \mp i \rho_2^-) = \left(e B (2 N+ 1) +  p_3^2  + m_\rho^2 \mp
2 e B \right) (\rho_1^- \mp i \rho_2^-)
\end{equation}
with $p_3$ the momentum of the meson in the $x_3$-direction.
The combinations $(\rho_1^- - i \rho_2^-)$ and $(\rho_1^+ + i \rho_2^+)$
have spin $s_3 = 1$ parallel to the external magnetic field. In the lowest energy state  ($N=0$, $p_3=0$) their effective mass,
\begin{equation} \label{effmass}
m_{eff}^2 = m_\rho^2 - eB,
\end{equation}
can become zero if the magnetic field is strong enough.

This suggests that the QCD vacuum in a strong magnetic field $B = F_{12}$ is
unstable towards condensation of the fields
\[
\rho = \rho_1^- - i \rho_2^- \quad \mbox{and} \quad \rho^\dagger = \rho_1^+
+ i \rho_2^+
\]
at a critical value of the magnetic field
\begin{equation}
eB_{c} = m_\rho^2.
\end{equation}
Here, in our first approximation, the $B$-dependence of $m_\rho^2$ itself was ignored (as in \cite{Chernodub:2010qx}).
At $B \approx B_c$ the decay modes of rho mesons into pions, which normally would guarantee a
short lifetime of the rho mesons, are no longer kinematically allowed because the charged pions' masses increase due to the magnetic field (see discussion in \cite{Chernodub:2010qx}).

\begin{figure}[h!]
  \hfill
  \begin{minipage}[t]{.45\textwidth}
    \begin{center}
      \scalebox{0.8}{
  \includegraphics{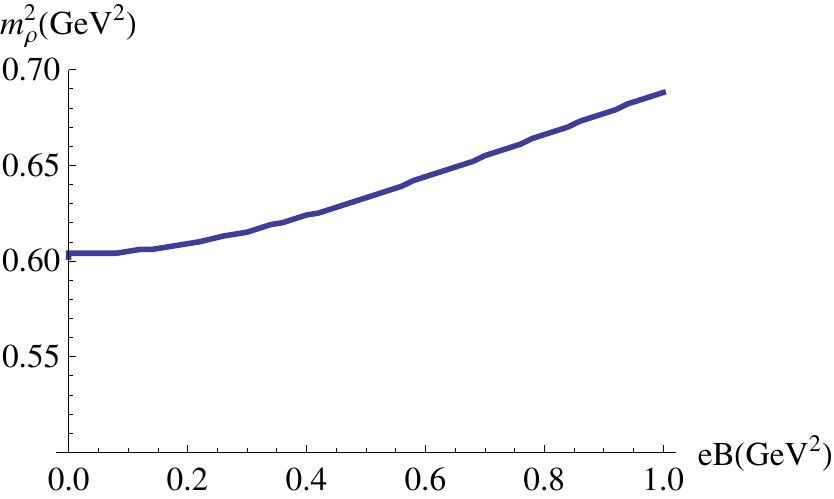}}
      \caption{Mass of the rho meson squared as a function of the magnetic
field.}
      \label{mrho}
    \end{center}
  \end{minipage}
  \hfill
  \begin{minipage}[t]{.45\textwidth}
    \begin{center}
      \scalebox{0.8}{
  \includegraphics{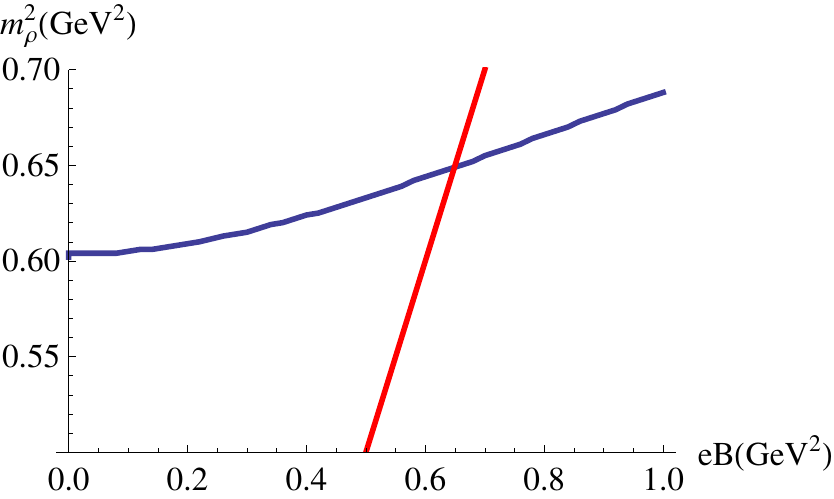}}
      \caption{$eB_{c}$ is determined as the value where the effective rho meson mass  $m_{eff}^2 = m_\rho^2 - eB$ becomes zero, i.e.~at the crossing of the 2 curves.}  
      \label{mrho_en_B}
    \end{center}
  \end{minipage}
  \hfill
\end{figure}
Next, we can calculate the $B$-dependence of $m_\rho^2$ in (\ref{landau})
itself, the lowest eigenvalue of the equation (\ref{eigwvgl}),
which depends on $B$ both explicitly and implicitly through the changed
embedding of the probe branes, represented by the value of $u_0(B)$.
This reflects the effect of chiral magnetic catalysis.
The result of our numerical study \cite{nele} is depicted in Figure \ref{mrho},
showing that $m_\rho^2$ is an increasing function of $B$, as expected.
The value of the magnetic field at the onset of the condensation becomes
slightly larger:
\begin{equation}
eB_{c} = 1.08 m_\rho^2.
\end{equation}
We clearly observe that the effect is of the order of a few percent, which
is consistent with the rough estimate provided in \cite{Chernodub:2010qx}.

\section{Split between deconfinement and chiral symmetry restoration
temperature}

\begin{figure}[h!]
  \centering
  \scalebox{0.8}{
  \includegraphics{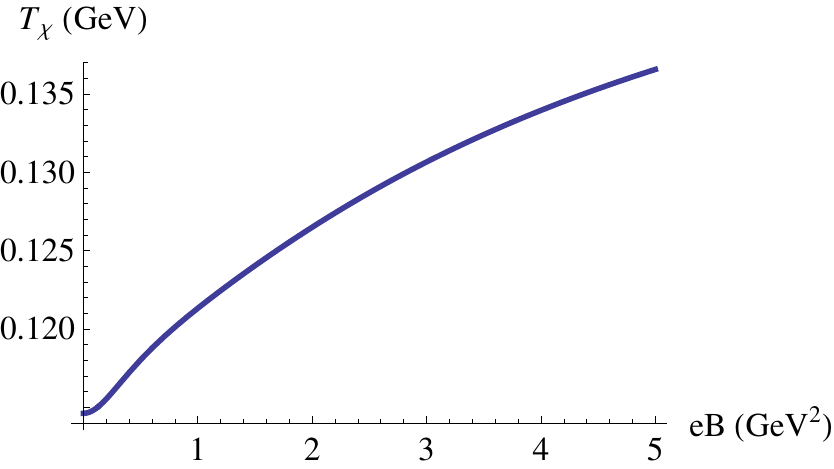}}
  \caption{The chiral symmetry restoration temperature as a function of the
magnetic field. The deconfinement temperature remains at $T_c\approx 0.115$
GeV.}\label{Tchiral}
\end{figure}

In the Sakai-Sugimoto model the chiral symmetry restoration temperature
$T_\chi$ is known to increase when a background magnetic field is present
\cite{Johnson:2008vna}. Because of the way the magnetic field is introduced
in the Sakai-Sugimoto model, the deconfinement temperature $T_c$ is
independent\footnote{$T_c$ is determined solely from the background
geometry, which is totally $B$-independent in the current approach.} of $B$,
with $T_c\approx 115 \textrm{ MeV}$ for our numerical values of the
holographic parameters \cite{nele}, so there will arise a split between the
deconfinement transition and chiral symmetry restoration. We numerically
calculate this split, assuming $T_\chi(B=0) = T_c$, and find it to be  $1.2
\%$ at $B = 15 m_\pi^2\approx0.28$ GeV$^2$ and $3.2 \%$ at $B = 30
m_\pi^2 \approx 0.57$ GeV$^2$, see Figure \ref{Tchiral}. The percent
estimation of the magnitude of the  split is in quantitative agreement with
\cite{Gatto:2010pt,D'Elia:2010nq}, and in qualitative agreement with
\cite{Mizher:2010zb,Gatto:2010qs}. Notice however that none of these papers
work in the chiral limit, and that all of these also witnessed an
increasing $T_c$ in terms of $B$. Results in the chiral limit can be
extracted from the work \cite{Fukushima:2010fe} wherein, interestingly,
$T_c$ seems to depend only marginally on $B$, at least up to $eB=20
m_\pi^2$, the maximum value considered there.
It is perhaps interesting to notice that for ever-growing values of $B$, a
saturation (horizontal asymptote) is seen in all our quantities ($u_0(B)$,
$m_q(B)$ and $T_\chi(B)$), in agreement with what was found in
\cite{Johnson:2008vna}. It is unclear whether this behaviour really
corresponds to QCD. From QCD-related models like
\cite{Mizher:2010zb,Gatto:2010pt,Gatto:2010qs,Fraga:2008qn} and lattice
simulations like \cite{Buividovich:2008wf}, one rather expects that
e.g.~$m_q(B)$ (or more precisely, the chiral condensate) and $T_\chi(B)$
will keep growing for large $B$. It looks tempting to associate such a
discrepancy to the probe approximation. At large values of $B$, perhaps the
quark back reaction might become more prominent, leading to other results
than found for e.g.~$m_q(B)$ at large $B$. Likewise, also the deconfinement
temperature might become $B$-dependent beyond the probe approximation, in
harmony with works like
\cite{Mizher:2010zb,Gatto:2010pt,Gatto:2010qs,Fraga:2008qn}. It thus
seems interesting to study the back reaction of the flavour branes on the
background geometry, with or without magnetic field present\footnote{In
absence of the probe branes, a constant magnetic field $B$ does not change
the background geometry, see \cite{Filev:2007gb} and references therein.}.

So far, we did not study the actual condensation of the rho mesons, a work
which we relegate to the future. Let us however already refer to
\cite{erdmenger} for related work using a different holographic setup, and to \cite{Preis:2010cq} where the splitting of the 2 transitions was also studied in the Sakai-Sugimoto model. However, in contrast to our analysis, no results in physical units were presented, which makes explicit comparison with QCD, or QCD models, rather difficult. We did determine all ``unphysical'' parameters of the Sakai-Sugimoto model, like $L$ or $M_K$, in such a way that known QCD physical constants are reproduced as good as possible. It would also be interesting to include the pions in a future analysis. Since
the rotational symmetry is broken by $\vec{B}=B\vec{e}_3$, one might expect
a mixing between the longitudinal vector mesons and pions in some manner. It
is apparent that such happens when taking into account the to the chiral
anomaly related Chern-Simons piece of the action \cite{nele}, but as said
before, this effect is suppressed by an additional inverse 't Hooft
coupling. For sure, getting a clear holographic understanding of the occurrence of a
magnetically induced superconducting QCD vacuum at zero temperature, as
touched upon in \cite{Chernodub:2010qx,Chernodub:2011mc}, will
need further efforts.

\section*{Acknowledgments}

\noindent We are grateful to M.~Chernodub, A.~J.~Mizher and D.~Vercauteren
for helpful discussions. We would also like to thank all the participants of
the workshop ``The many faces of QCD'' for a pleasant atmosphere.
N.~Callebaut and D.~Dudal and  are  supported by the Research Foundation
Flanders (FWO).

\end{document}